# Annealing-induced enhancement of ferromagnetism and nano-particle formation in ferromagnetic-semiconductor GeFe

Yuki Wakabayashi, Yoshisuke Ban, Shinobu Ohya, and Masaaki Tanaka

*Department of Electrical Engineering and Information Systems, The University of Tokyo, 7-3-1 Hongo, Bunkyo-ku, Tokyo 113-8656, Japan*

Abstract

We report the annealing-induced enhancement of ferromagnetism and nano-particle formation in group-IV-based ferromagnetic-semiconductor GeFe. We successfully increase the Curie temperature of the $Ge_{0.895}Fe_{0.105}$ film up to ~220 K while keeping a single ferromagnetic phase when the annealing temperature is lower than 500°C. In contrast, when annealed at 600°C, single-crystal GeFe nano-particles with stacking faults and twins, which have a high Curie temperature nearly up to room temperature, are formed in the film. Our results show that annealing is quite effective to improve the magnetic properties of GeFe for high-temperature-operating spin-injection devices based on Si or Ge.

The ferromagnetic materials based on group-IV-semiconductor Ge, including Ge-based ferromagnetic semiconductors (FMSs) and ferromagnetic nano-particles embedded in Ge, are very promising for future Si-based spintronic devices. Among them, GeMn and GeFe have been intensively studied. They can be epitaxially grown on Si substrates,[1,2] flat and smooth interfaces with Si without a disordered interfacial layer can be formed, and there is no conductivity mismatch problem with Si. Therefore, they will be efficient spin injectors and detectors which can suppress the spin flip scattering across the interfaces. Many studies have been carried out on Mn-doped Ge ($Ge_{1-x}Mn_x$) films,[3-6] and they frequently contain ferromagnetic intermetallic precipitates such as $Mn_5Ge_3$, $Mn_{11}Ge_8$[5] or amorphous $Ge_{1-x}Mn_x$.[6] In the case of $Ge_{1-x}Fe_x$ films, we can grow single-crystal films of a diamond type with the non-uniform distribution of Fe atoms without any intermetallic Fe-Ge compounds, which is appropriate for heterostructure applications, but the current problem of GeFe is its low Curie temperature ($T_C$) which is at the highest 170 K so far.[7-9] Recently, however, GeFe quantum dots with a high $T_C$ of ~400 K without any observable precipitates have been reported.[10] Thus, if we can grow GeFe quantum dots (or nano-particles) inside a Ge film with a flat surface or interfaces with other layers (or substrates), they are very promising. In fact, the ferromagnetic MnAs nano-particles embedded in GaAs have shown intriguing properties induced by Coulomb blockade and spin-dependent tunneling.[11,12] In III-V-based FMS (Ga,Mn)As, post-growth annealing is known to be a powerful technique to improve the $T_C$ by removing the interstitial Mn atoms from the (Ga,Mn)As layer.[13,14] Here, we investigate the annealing effect on GeFe in order to enhance the ferromagnetism of GeFe.

The $Ge_{0.895}Fe_{0.105}$ thin film studied here was epitaxially grown on a Ge(001) substrate by low-temperature molecular beam epitaxy (LT-MBE). The growth process is described as follows. After the Ge(001) substrate was chemically cleaned and its surface was hydrogen-terminated by buffered HF solution, it was introduced into the MBE growth chamber through an oil-free load-rock system. After degassing the substrate at 400°C for 30 min and successive thermal cleaning at 900°C for 15 min, a 30-nm-thick Ge buffer layer was grown at 200°C, followed by the growth of a 60-nm-thick $Ge_{0.895}Fe_{0.105}$ layer at 240°C. The *in-situ* reflection high-energy electron diffraction (RHEED) was used to observe the crystallinity and morphology of the surface during the growth. The diffraction pattern of the Ge buffer layer surface showed intense and sharp 2 × 2 streaks, and the $Ge_{0.895}Fe_{0.105}$ surface also showed a 2 × 2 pattern with no extra spots, indicating two-dimensional epitaxial growth. Post-growth annealing was carried out in nitrogen atmosphere for 30 min at 400, 500, and 600°C.

The crystallographic analyses of the $Ge_{0.895}Fe_{0.105}$ films were performed by high-resolution transmission electron microscopy (HRTEM). Figure 1 (a) and (b) show the HRTEM lattice images of the $Ge_{0.895}Fe_{0.105}$ film (a) as grown and (b) annealed at 500°C projected along the Ge [110] axis. The both images indicate that the $Ge_{0.895}Fe_{0.105}$ layers have a diamond-type single-crystal structure with an atomically flat surface. As discussed in Ref. 9, the color (black and white) contrast in the GeFe layer comes from the non-uniform distribution of Fe atoms and stacking-fault defects, but there are no other ferromagnetic intermetallic Fe-Ge precipitates with a different crystal structure. Figure 1 (c) shows a HRTEM lattice image of the $Ge_{0.895}Fe_{0.105}$ film annealed at 600°C projected along the Ge [110] axis, where we see many nano-particles are formed in the film. Figure 2 (a) shows the magnified view of the HRTEM lattice image of one of the nano-particles, indicating that the nano-particles contain periodic twins and stacking faults. Spatially resolved transmission electron diffraction (TED) combined with energy-dispersive X-ray spectroscopy (EDX) was measured to analyze the nano-particles in detail. The EDX observations revealed the local Fe compositions of the $Ge_{0.895}Fe_{0.105}$ film annealed at 500°C and 600°C. In the GeFe film annealed at 500°C, the local Fe concentration at *1 (white contrast region) and *2 (black contrast region) in Fig.1 (b) were approximately 8 and 23%, respectively. In the GeFe film annealed at 600°C, the local Fe concentration at *1 (homogeneous diamond crystal structure region) and *2 (inside the nano-particle) in Fig. 2 (a) were approximately 5 and 25%, respectively. These results indicate that the higher the annealing temperature is, the larger the non-uniformity of the Fe concentration becomes. Figure 2 (b) shows the TED image at *1, exhibiting the diffraction pattern of the diamond-type lattice structure with weak extra spots due to stacking-fault defects. Figure 2 (c) shows the TED image at *2, indicating a similar diffraction pattern of the diamond-type lattice structure including clear twins and stacking faults.[15] In either TED images, we do not see any diffractions from precipitates with crystalline Fe-Ge intermetallic compounds of other crystal structures.

Magneto-optical measurements were performed in order to investigate the magnetic properties of the $Ge_{0.895}Fe_{0.105}$ films. Magnetic circular dichroism (MCD), which is defined by the difference between the optical reflectances of right- and left-circular polarized lights, is an effective tool to identify what is the origin of the ferromagnetism inside the film. This is because the MCD intensity is enhanced at the critical-point energies of the band structure due to the spin splitting caused by the s,p-d exchange interactions, which are considered to be the origin of the ferromagnetism in FMSs,[16] and because it is proportional to the vertical component of the magnetization ($M$). Figure 3 (a) shows the MCD spectra of the $Ge_{0.895}Fe_{0.105}$ films as grown and annealed at 400, 500, and 600°C (from the top to the bottom) with a magnetic field of 1 T applied perpendicular to the film plane at 10 K. The normalized MCD spectra of the $Ge_{0.895}Fe_{0.105}$ film annealed at 500°C at different magnetic fields (0.2 T, 0.5 T, and 1 T) at 10 K are superimposed on a single spectrum over the whole photon energy range (See Fig. S1 in Supplemental material.),[17] indicating that the MCD spectrum comes from the single FMS phase of GeFe[18] even though it has the non-uniform distribution of Fe atoms. This suggests that the locally high-Fe-concentration region and low-Fe-concentration region are magnetically coupled by the s,p-d exchange interaction, which results in the magnetically homogeneous behavior. All the samples show the $E_1$ peak at around 2.3 eV corresponding to the L point of bulk Ge and a very broad peak ($E^*$) at around 1.5 eV. The $E_1$ peak enhanced by the s,p-d exchange interaction is a characteristic property of ferromagnetic semiconductors.[19] The origin of the $E^*$ peak, which tends to be enhanced as the annealing temperature is increased, is not clear, but there are two possibilities; optical transitions from impurity bands, which have been observed in III-V-based FMS $Ga_{1-x}Mn_xAs$,[20] and d-d transitions related to crystal-field splitting of substitutional Fe atoms. We think that the annealing increases the density of substitutional or effective Fe atoms. This enhances the density of states of the impurity bands or d states, leading to the enhancement of the $E^*$ peak. Meanwhile, the $E_1$ peak is suppressed by the annealing at 600°C, which is related to the phase separation shown in Fig. 1 (b). Generally, a MCD spectrum of a phase-separated material is expressed by a sum of the ones of these phases. During the annealing at 600°C, the Fe atoms are removed from the region *1 and it becomes nearly pure Ge, which results in suppressing the $E_1$ peak. Figures 3 (b)-(d) show the magnetic-field ($B$) dependence of the normalized MCD intensities at the $E^*$ (1.5 eV, solid curve) and $E_1$ (2.3 eV, dotted curve) in (b) the as-grown $Ge_{0.895}Fe_{0.105}$ film and in the $Ge_{0.895}Fe_{0.105}$ film annealed at (c) 500°C and (d) 600°C. Figure 3 (c) also shows the $B$ dependence of the normalized

-$M$ of the Ge$_{0.895}$Fe$_{0.105}$ film annealed at 500°C (green curve) measured by superconducting quantum interference device (SQUID). Here the diamagnetic signal of the Ge substrate was subtracted from the raw $M$ data. In the as-grown GeFe film and annealed film at 500°C, the shapes of the curves at 1.5 and 2.3 eV are identical with each other, which means that the $E^*$ and $E_1$ peaks come from the single ferromagnetic phase of FMS GeFe as previously mentioned. Moreover, Fig. 3 (c) shows that the hysteresis loops of MCD have the same shape as that of the $M$-$B$ curve measured by SQUID at 10 K in the Ge$_{0.895}$Fe$_{0.105}$ film annealed at 500°C. This indicates that the $M$ data measured by SQUID has the same origin as that induces the spin splitting of the energy band of GeFe. Therefore, We conclude that the origin of the ferromagnetism is only the single FMS phase of GeFe[21]. In contrast, after the annealing at 600°C, the curves are not identical, which indicates that there are two or more ferromagnetic phases in the film. These results mean that the single FMS phase of GeFe film was phase-separated magnetically as well as crystallographically by the annealing at 600°C.

Figure 4 shows the Curie temperature ($T_C$) estimated from the Arrott plots of the MCD-$B$ curves measured at 1.5 eV (square) and 2.3 eV (triangle). Note that the clear hysteresis loops are seen up to $T_C$ in the MCD-$B$ curves (See Fig. S2-S4 in Supplemental material.).[17] In the as-grown film, the $T_C$ values at both of the photon energies are the same. When annealed at 400 or 500°C, even though the film is magnetically homogeneous as discussed above, we see a slight difference of the $T_C$ values between at 1.5 and 2.3 eV, probably due to the non-uniformity of the Fe concentration which is enhanced by the annealing. When annealed at 600°C, the $T_C$ values at 1.5 and 2.3 eV are completely different, reflecting the phase separation.

The magnetic properties of the films were measured by SQUID. Figures 5 (a)-(d) show the magnetization versus temperature ($M$-$T$) curves of the Ge$_{0.895}$Fe$_{0.105}$ samples as grown and annealed at 400, 500, and 600°C, respectively. Here, two processes of field cooling (FC, red curve) and zero field cooling (ZFC, blue curve) were used with a magnetic field of 100 Oe applied perpendicular to the film plane. These $M$-$T$ curves in these figures are characterized by the irreversibility between the ZFC and FC processes. The $M$-$T$ curves measured by SQUID have a long tail above the $T_C$ values estimated by the Arrott plots of the MCD-$B$ curves obtained at 1.5 (red arrows) and 2.3 eV (blue arrows), which means that the ferromagnetic interactions between the Fe spins in the locally high-Fe-concentration regions in the GeFe films are still remaining above the $T_C$ values estimated above. In all the samples, a cusp corresponding to the blocking temperature ($T_B$) is seen at around 15 K in the ZFC curves, being a characteristic feature of a magnetic random system like spin glass. This randomness does not come from the ferromagnetic intermetallic precipitates but from the non-uniform distribution of Fe atoms. The same phenomena were observed in II-VI-based FMS ZnCrTe[22] and Ni$_3$Mn.[23] The notable point is that, when annealed at 600°C, the sample shows an another cusp at 260 K, which indicates the occurrence of phase separation and superparamagnetism. Moreover, the magnetic moment persists up to room temperature, meaning that the nano-particles, which have the high Fe concentration with the periodic twins and stacking faults, have a high $T_C$ up to room temperature. These results obtained by SQUID show that the magnetic phase separation occurs at 600°C, being consistent with the crystallographic and MCD analyses mentioned above.

The $T_C$ values of the well-known equilibrium Fe-Ge compound phases are much higher than room temperature,[24-26] which suggests that the ferromagnetic nano-particles obtained in this study are a new magnetic phase which has a diamond crystal structure with periodic twins and stacking faults. The high $T_C$ of the nano particles may come from the high Fe concentration, quantum confinement, or band structure modulation associated with formation of twins.[27] Twin boundaries are viewed to be a regional wurtzite structure so these nano-particles, which include periodic twin boundaries, may change their band structure and physical properties.

In summary, annealing of the GeFe thin film in nitrogen atmosphere was shown to be quite effective to enhance the ferromagnetism of GeFe. When the annealing temperature is lower than 500°C, $T_C$ is increased up to ~220 K while keeping a single FMS phase. When annealed at 600°C, ferromagnetic nano-particles with a high $T_C$ up to room temperature, which have a diamond crystal structure with twins and stacking-faults, are formed in the GeFe film. Both types of films have a relatively flat surface, and thus they are very promising for realizing Si-based spin devices.

This work was partly supported by Giant-in-Aids for Scientific Research including Specially Promoted Research, Project for Developing Innovation Systems of MEXT, and FIRST program of JSPS. This work was partly conducted in Research Hub for Advanced Nano Characterization, The University of Tokyo, supported by the MEXT, Japan.

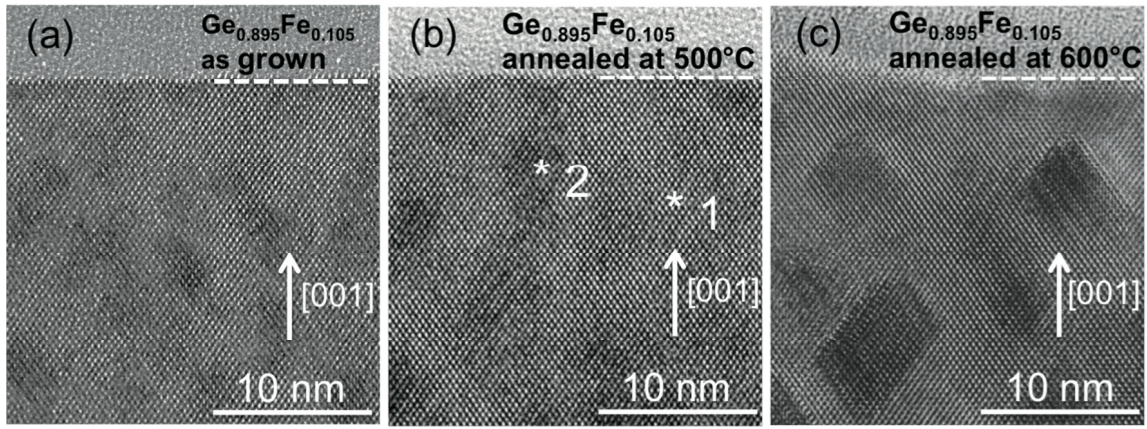

Fig. 1 High resolution transmission electron microscopy (HRTEM) lattice image projected along the Ge[110] axis of $Ge_{0.895}Fe_{0.105}$ (a) as grown and after annealing at (b) 500°C and (c) 600°C.

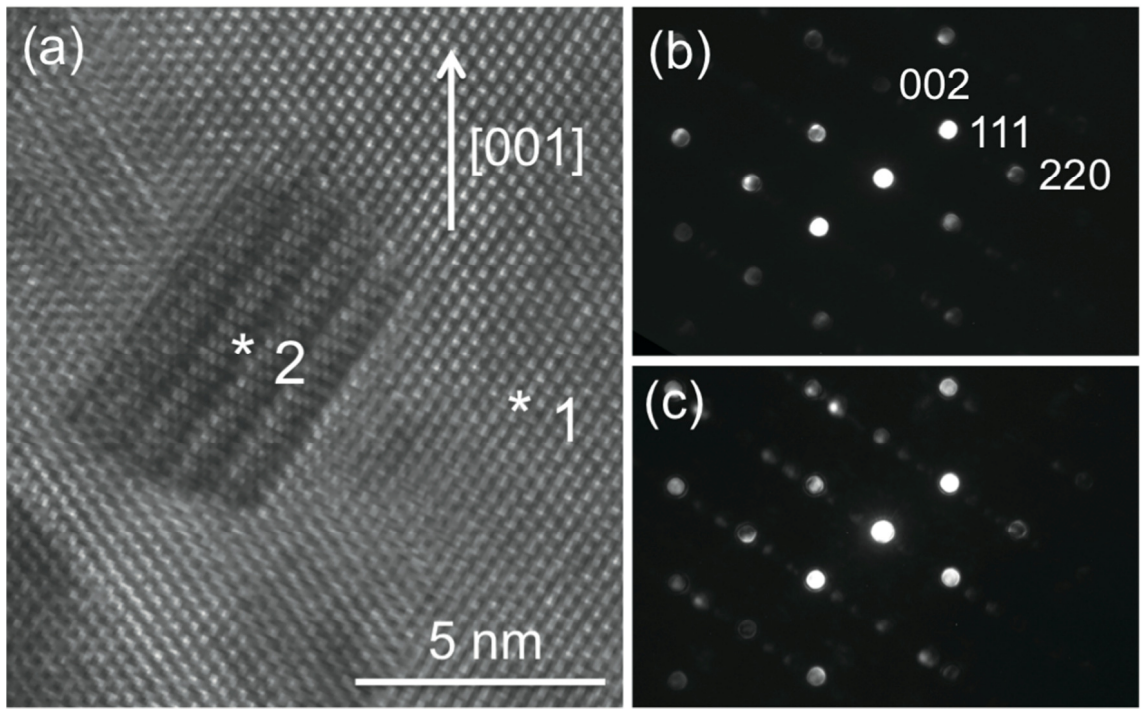

Fig. 2 (a) HRTEM lattice image projected along the Ge[110] axis of a ferromagnetic nano-particle formed in the $Ge_{0.895}Fe_{0.105}$ film after annealing at 600°C. (b) (c) TED images taken in (b) the diamond crystal structure region (*1) and in (c) the nano-particle (*2).

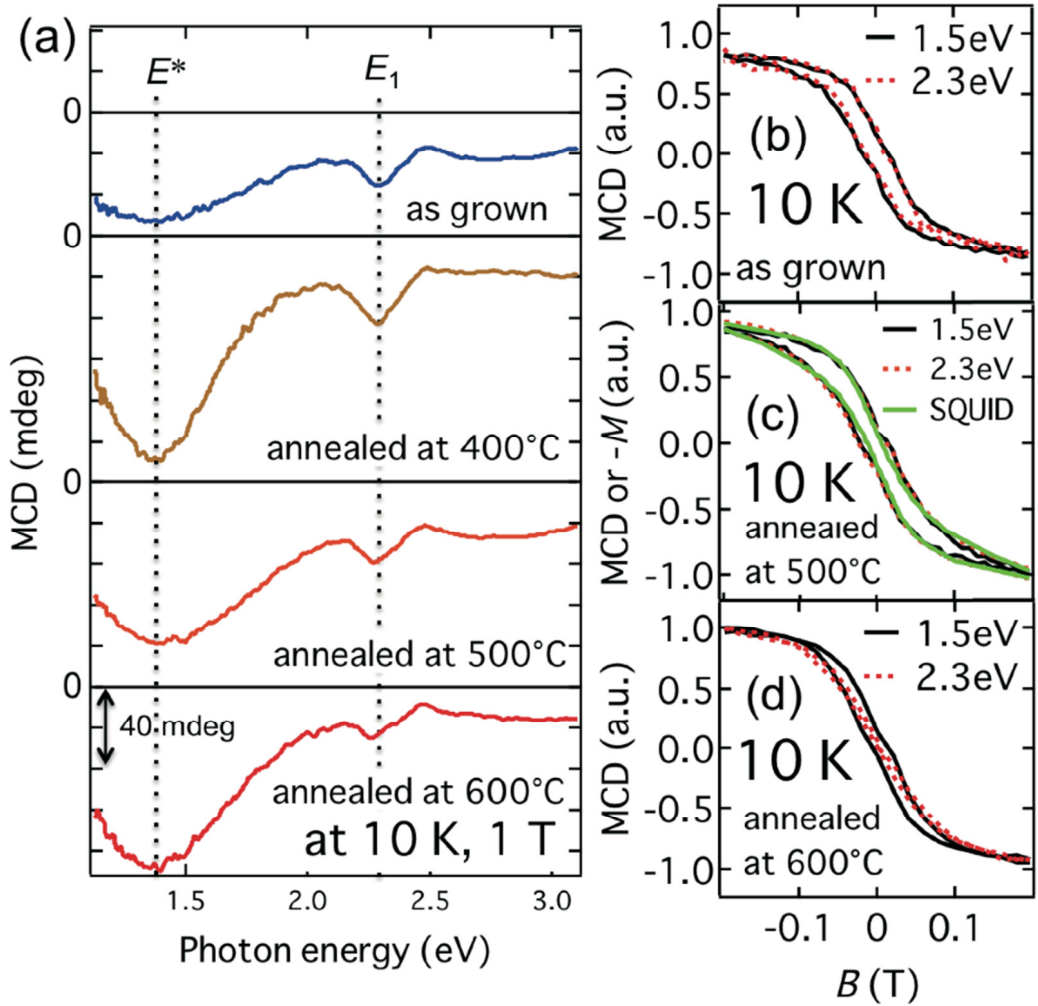

Fig. 3 (a) MCD spectra of $Ge_{0.895}Fe_{0.105}$ films as grown and annealed at 400, 500, and 600°C (from the top to the bottom) with a magnetic field of 1 T applied perpendicular to the film plane at 10 K. (b)-(d) Magnetic-field dependence of the MCD at 1.5 eV ($E^*$ peak) and 2.3 eV ($E_1$ peak) in the GeFe film (b) as grown and annealed at (c) 500°C and (d) 600°C. In (c), the green curve expresses the magnetic-field dependence of the magnetization measured by SQUID in the $Ge_{0.895}Fe_{0.105}$ film annealed at 500°C.

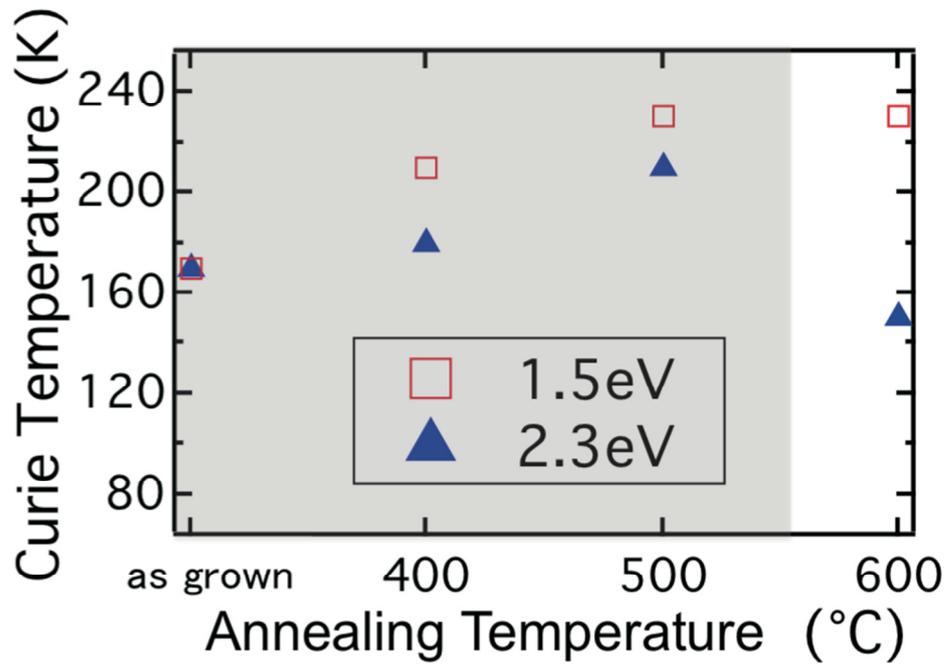

Fig. 4 Curie temperature as a function of the annealing temperature of the $Ge_{0.895}Fe_{0.105}$ films estimated by the Arrott plot ($MCD^2 - B/MCD$) at the photon energies of 1.5 eV (square) and 2.3 eV (triangle), where $B$ is the magnetic field.

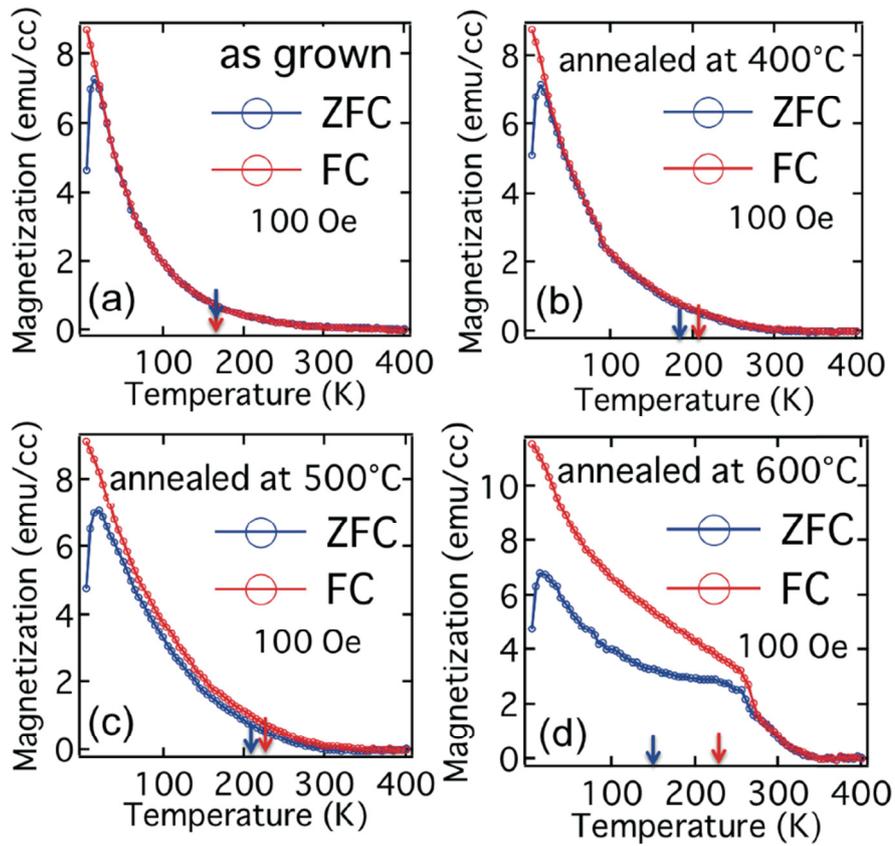

Fig. 5 Magnetization versus temperature (*M-T*) curves of the Ge$_{0.895}$Fe$_{0.105}$ samples (a) as grown and annealed at (b) 400, (c) 500, and (d) 600°C. The measurements were performed in the two processes of field cooling (FC, red curve) and zero field cooling (ZFC, blue curve) with a magnetic field of 100 Oe applied perpendicular to the film plane. The red and blue arrows are the $T_C$ values estimated by the Arrott plots of the MCD-*B* curves obtained at 1.5 and 2.3 eV, respectively.

**Supplemental Material for Annealing-induced enhancement of ferromagnetism and nano-particle formation in ferromagnetic-semiconductor GeFe**

Yuki Wakabayashi, Yoshisuke Ban, Shinobu Ohya, and Masaaki Tanaka
*Department of Electrical Engineering and Information Systems, The University of Tokyo, 7-3-1 Hongo, Bunkyo-ku, Tokyo 113-8656, Japan*

Figure S1 shows the normalized MCD spectra of the $Ge_{0.895}Fe_{0.105}$ film annealed at 500°C with various magnetic fields applied perpendicular to the film plane at 10 K. Figure S2 (S3) shows the Arrott plots of the MCD-*B* curves at 1.5 (2.3) eV measured at various temperatures for (a) the as-grown film and annealed films at (b) 400, (c) 500, and (d) 600°C. Figure S4 shows the magnetic-field dependence of MCD at $E_1$(2.3eV) in the $Ge_{0.895}Fe_{0.105}$ film annealed at 500°C, measured at various temperatures from 10 K to 240 K.

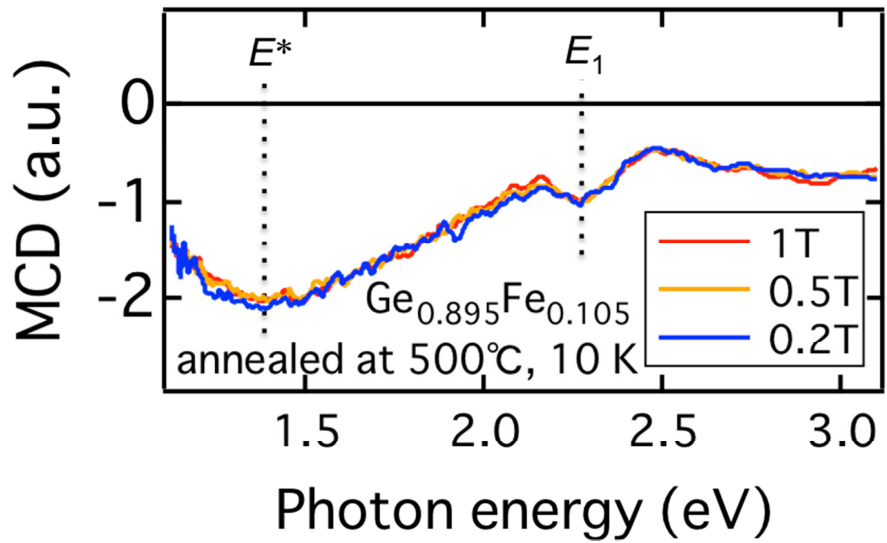

Fig. S1. Normalized MCD spectra of the $Ge_{0.895}Fe_{0.105}$ film annealed at 500°C with magnetic fields of 1 T (red curve), 0.5 T (orange curve), and 0.2 T (blue curve) applied perpendicular to the film plane at 10 K.

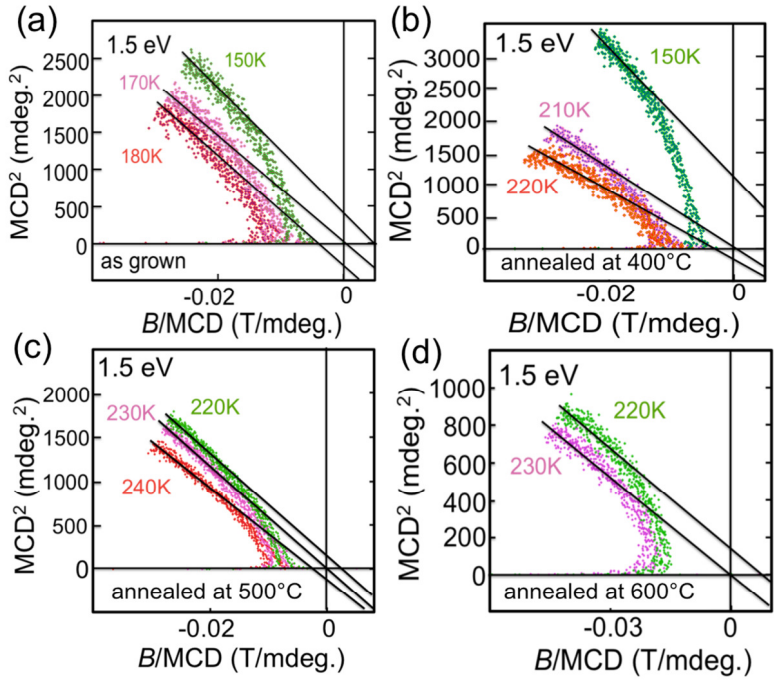

Fig. S2. Arrott plots of the MCD-*B* curves at 1.5 eV of $Ge_{0.895}Fe_{0.105}$ measured at various temperatures for (a) the as-grown film and the films annealed at (b) 400, (c) 500, and (d) 600°C.

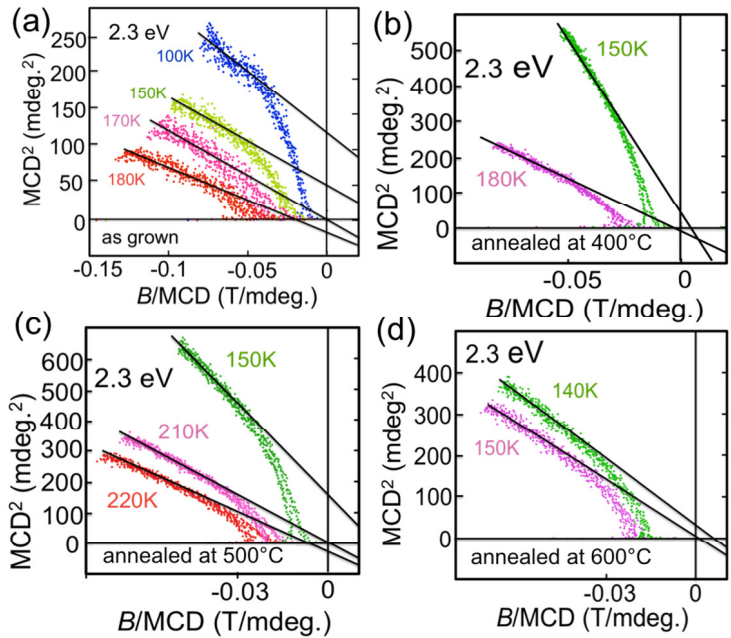

Fig. S3. Arrott plots of the MCD-*B* curves at 2.3 eV of $Ge_{0.895}Fe_{0.105}$ measured at various temperatures for (a) the as-grown film and the films annealed at (b) 400, (c) 500, and (d) 600°C.

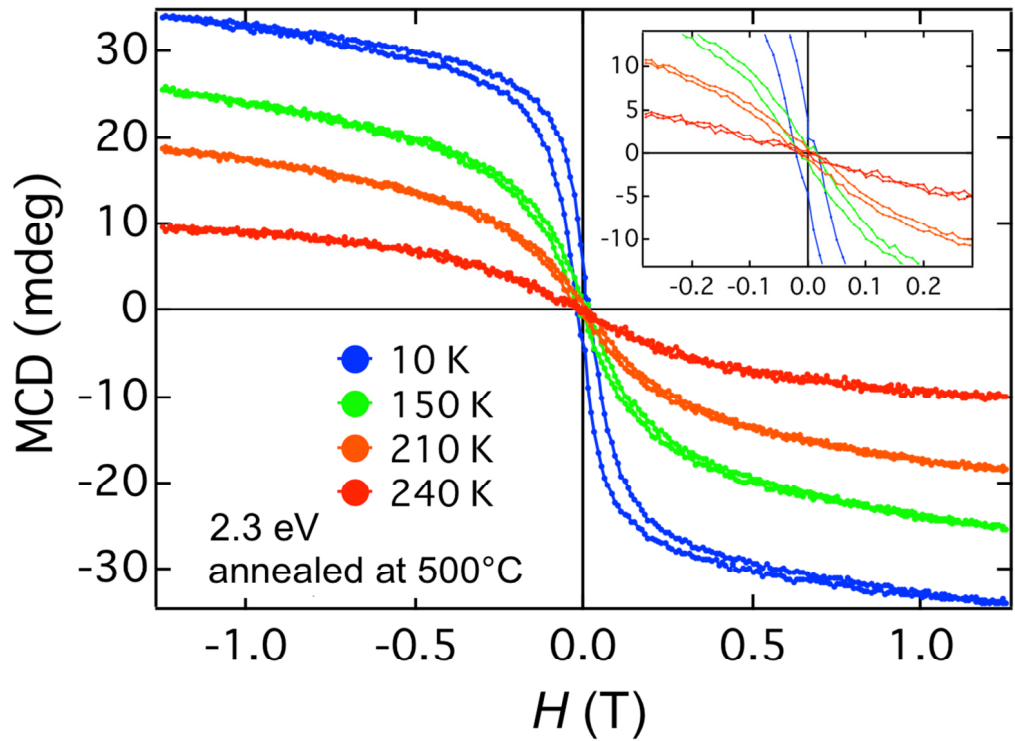

Fig. S4. Magnetic-field dependence of MCD at $E_1$ (2.3eV) of the $Ge_{0.895}Fe_{0.105}$ film annealed at 500°C measured at 10 K (blue curve), 150 K (green curve), 210 K (orange curve), and 240 K (red curve). The inset shows the close-up view near zero magnetic field.